\documentclass[sigconf, nonacm]{acmart}
\newcommand\vldbdoi{XX.XX/XXX.XX}
\newcommand\vldbpages{XXX-XXX}
\newcommand\vldbvolume{14}
\newcommand\vldbissue{1}
\newcommand\vldbyear{2020}
\newcommand\vldbauthors{\authors}
\newcommand\vldbtitle{\shorttitle}
\usepackage{enumitem}
\usepackage{multicol}
\usepackage{url}
\usepackage{graphicx}

\begin{document}
\title{A Survey on Advancing the DBMS Query Optimizer: Cardinality Estimation, Cost Model, and Plan Enumeration}
\author{Hai Lan}
\affiliation{%
  \institution{RMIT University}
  \city{Melbourne}
  \state{Australia}
}
\email{hai.lan@rmit.edu.au}

\author{Zhifeng Bao}
\affiliation{%
  \institution{RMIT University}
  \city{Melbourne}
  \country{Australia}
}
\email{zhifeng.bao@rmit.edu.au}

\author{Yuwei Peng*}
\affiliation{%
  \institution{Wuhan University}
  \city{Wuhan}
  \country{China}
}
\email{ywpeng@whu.edu.cn}

\begin{abstract}
  Query optimizer is at the heart of the database systems. 
  Cost-based optimizer studied in this paper is adopted in almost all current database systems.
  A cost-based optimizer introduces a plan enumeration algorithm to find a (sub)plan, 
  and then uses a cost model to obtain the cost of that plan, and selects the plan with the lowest cost.
  In the cost model, cardinality, the number of tuples through an operator, plays a crucial role.
  Due to the inaccuracy in cardinality estimation, errors in cost model, and the huge plan space, 
  the optimizer cannot find the optimal execution plan for a complex query in a reasonable time. 
  In this paper, we first deeply study the causes behind the limitations above. 
  Next, we review the techniques used to improve the
  quality of the three key components in the cost-based optimizer, cardinality estimation, cost model, and plan enumeration.
  We also provide our insights on the future directions for each of the above aspects.
  
  \end{abstract}

\maketitle

\section{Introduction}\label{introduction}
Query optimizer is at the heart of relational database management systems (RDBMSes) and some big data process engines, e.g. SCOPE~\cite{SCOPE_ChaikenJLRSWZ08}.
Given a query written in a declarative language (e.g., SQL), the optimizer finds the most efficient execution plan (also called physical plan) and feeds it to the executor.
Thus, most of the time, the users only think over how to transform their requirements to a valid query without the need to analyze how to run the query efficiently. 
 Almost all systems adopt a cost-based optimizer based on the architecture of System R~\cite{JOIN_T_SelingerACLP79} or Volcano/Cascades~\cite{Volcano93,Cascades95}.

\begin{figure}[h]
    \centering
    \includegraphics[width=\linewidth]{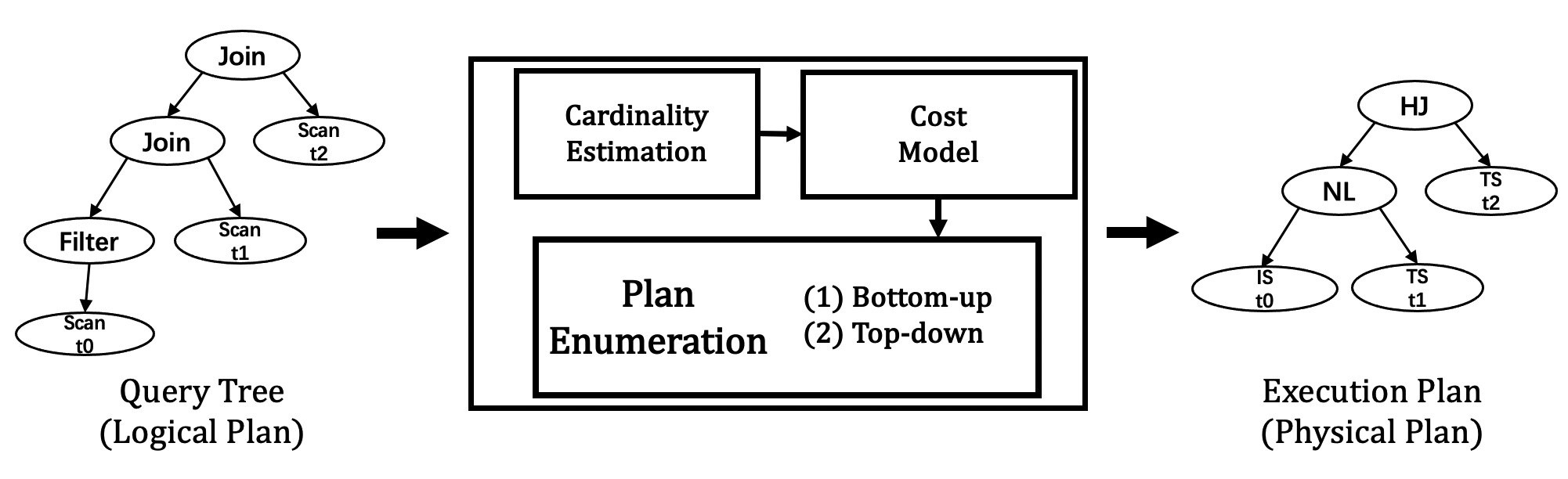}
    \caption{Query optimizer architecture. IS, HJ, NL, and TS refer to index scan, hash join, nestloop join, and table scan.}
    \Description{Cost-based optimizer.}
    \label{op_arch}
  \end{figure}

 Figure \ref{op_arch} illustrates the three most important components in a cost-based optimizer: \textbf{cardinality estimation (CE)}, \textbf{cost model (CM)}, and \textbf{plan enumeration (PE)}. 
 \textbf{CE} uses statistics of data and some assumptions 
 about data distribution, column correlation, and join relationship 
 to get the number of tuples generated by an intermediate operator\footnote{In some context, the cardinality in the database area refers to distinct count~\cite{CE_HarmouchN17}.}, which is also crucial
 for other search problems, e.g.,~\cite{DSE_sl1,DSE_sl2}. 
 \textbf{CM} can be regarded as a complex function that maps the current state of database and estimated cardinalities to the cost of executing a (sub)plan.
 \textbf{PE} is an algorithm to explore the space of semantically 
 equivalent join orders and find the optimal orders with minimal cost. 
 There are two principal approaches to find an optimal join
 order: bottom-up join enumeration via dynamic programming
 and top-down join enumeration through memorization. 

Theoretically, provided that the estimated cardinality and cost are accurate,  
and plan enumeration component can efficiently walk through the huge search space, this architecture can obtain the optimal execution plan in a reasonable time. However, it fails in reality.
Despite decades of work, cost-based query optimizers still make mistakes on “difficult”
queries due to the error in \textbf{CE}, the difficulty in building an accurate \textbf{CM}, and
the pain in finding the optimal \textbf{join orders} (\textbf{PE}) for complex queries. The details are presented in Section \ref{why_bad}, i.e., why the existing optimizer is still far from satisfaction.

There are lots of research studies proposed to improve the capability of the optimizer. In this paper, we present a survey on them. Specifically, we review the publications which are proposed to improve the
capabilities of the three key components in the optimizer, i.e., \textbf{CE}, \textbf{CM}, \textbf{PE}.

This paper makes the following contributions:
\begin{enumerate}[noitemsep]
  \item We summarize the reasons why the \textbf{CE}, \textbf{CM}, and \textbf{PE} do not perform well (Section \ref{why_bad}).
  \item We review the studies proposed to estimate cardinality more accurately. According to the techniques used, we categorize them into synopsis-based methods, 
  sampling-based methods, and learning-based methods (Section \ref{CE}).
  \item We review the work on improving the cost model. We classify them into three groups: improvement of the existing cost model, 
  cost model alternatives, and performance prediction for a single query (Section \ref{CM}).
  \item We review the techniques used in plan enumeration and study the non-learning methods used to handle large queries. 
  Besides, we review recent proposed methods, which adopt reinforcement learning to select the join order (Section \ref{PE}).
  \item In Sections \ref{CE}-\ref{PE}, we present our insights on the future directions respectively.
\end{enumerate}

There are two related surveys. In 1998, \citet{Chaudhuri98} reviews the work with non-learning methods on query optimizer. In the last two decades, many methods are proposed to improve the capability of the optimizer.
It is necessary to review the new work.
Recently, \citet{AI4DB} investigate how AI is introduced in the different parts of DBMS, such as monitoring, tuning, optimizer. In this paper, we focus on the query optimizer and give a 
comprehensive survey on the three key components of the optimizer. 
We summarize the learning-based and non-learning methods at the same time, review these work in details, and present possible future directions for each of them.

\section{Why Key Components In Optimizer Are Still Not Accurate?}\label{why_bad}
In this section, we summarize the reasons why the cardinality estimation, cost model, and plan enumeration do not perform well respectively.
The studies reviewed in this paper try to improve the quality of the optimizer by handling these shortages. 
\subsection{Cardinality Estimation}
Cardinality estimation is the ability to estimate the tuples generated by an operator and is used in the cost model to calculate the cost of that operator.
\citet{Lohman2014} points out that the cost model can introduce errors of at most 30\%, while the cardinality estimation can easily introduce errors of many orders of
magnitude. \citet{Leis2015} experimentally revisit the components, \textbf{CE}, \textbf{CM}, and \textbf{PE} in the classical
optimizers with complex workloads. They focus on the quality of the physical plan on multi-join queries and get the same conclusion with Lohman.

The errors in cardinality estimation are mainly introduced in three cases:
\begin{enumerate}[noitemsep]
    \item \textbf{Error in single table with predications.} Database systems usually take histograms as the approximate distribution 
    of data. Histograms are smaller than the original data. Thus, it cannot represent the true distribution entirely and some assumptions (e.g. uniformity
     on a single attribute, independence assumption among different attributes) are proposed. When those assumptions are not hold, estimation errors 
     occur, leading to sub-optimal plans. The correlation among attributes in a table is not unusual. Multi-histograms have been proposed. 
     However, it suffers from a large storage size.
    \item \textbf{Error in multi-join queries.} Correlations possibly exist in columns from different tables. However, there is no efficient way to get synopses
    between two or more tables. Inclusion principle has been introduced for this case. The cardinality of a join operator is calculated using the inclusion principle
   with cardinalities of its children. It has large errors when the assumption is not held.
   Besides, for a complex query with multiple tables, the estimation errors can propagate and amplify from the leaves to root of the plan.
   The optimizers of commercial and open-source database systems still struggle in cardinality estimation for mult-join queries~\cite{Leis2015}.
    \item \textbf{Error in user defined function.} Most of database systems support the user-defined function (UDF). When a UDF exists in the condition, there is no general method
    to estimate how many tuples satisfying it~\cite{Chaudhuri98}. 
\end{enumerate}

\subsection{Cost Model}\label{intro_cm}
Cost-based optimizers use a cost model to generate the estimate of cost for a (sub)query. The cost of (sub)plan 
is the sum of costs of all operators in it.

The cost of an operator depends on the hardware where the database is deployed, the operator's implementation, the number of tuples 
processed by the operator, and the current database state (e.g., data in the buffer, concurrent queries)~\cite{CM_ManegoldBK02}. Thus, a large number of magic numbers should be determined when combining
all factors, and errors in cardinality estimation also affect the quality of the cost model. Furthermore, when the cost-based optimizer is deployed in a distributed 
or parallel database, the cloud environment, or the cross-platform query engines, the complexity of cost model is increasing dramatically.
Moreover, even with the true cardinality, the cost estimation of a query is not linear to the running time, which may lead to a suboptimal execution plan~\cite{CM_L_KaoudiQCPTC20, CM_L_SiddiquiJQPL20}.
\subsection{Plan Enumeration}
Plan enumeration algorithm is used to find the optimal join order from the space of semantically 
equivalent join orders such that the query cost is minimized. It has been proven to be an NP-hard problem~\cite{NP84}.
Exhaustive query plan enumeration is a prohibitive task for large databases with multi-join queries. 
Thus, it is crucial to explore the right search space which should consist of the optimal join orders or approximately optimal join orders and design an efficient enumeration algorithm. 
The join trees in the search space could be zig-zag trees, left-deep trees, right-deep trees, and bushy trees or 
the subset of them. Different systems consider different forms of join tree. There are three enumeration algorithms in traditional 
database systems: (1) bottom up  join enumeration via dynamic programming (DP) (e.g., System R~\cite{JOIN_T_SelingerACLP79}), 
(2) top-down  join enumeration through memorization (e.g., Volcano/Cascades~\cite{Volcano93,Cascades95}), and (3) randomized algorithms (e.g., genetic algorithm in PostgreSQL~\cite{PG} with numerous tables joining).

Plan enumeration suffers from three limitations: (1) the errors in cardinality estimation and cost model, (2) the rules used to prune the search space, and (3) dealing with the queries with large number of tables.
When a query touches a large number of tables, optimizers have to sacrifice optimality and employ heuristics to keep optimization time reasonable, like genetic algorithm in PostgreSQL, greedy method in DB2, which
usually generates poor plans.

We should notice the errors in cardinality will propagate to the cost model and lead to suboptimal join order. Eliminating or reducing the errors in 
cardinality is the first step to build a capable optimizer as Lohman~\cite{Lohman2014} says \textit{“The root of all evil, the Achilles Heel of query optimization, is the estimation of the size of intermediate results, known as cardinalities”}.

In the following three sections, we summarize the research efforts made to handle limitations in CE, CM, and PE, i.e, how to make the query optimizer good.

\section{Cardinality Estimation}\label{CE}
\begin{figure}[t]
    \centering
    \includegraphics[width=\linewidth]{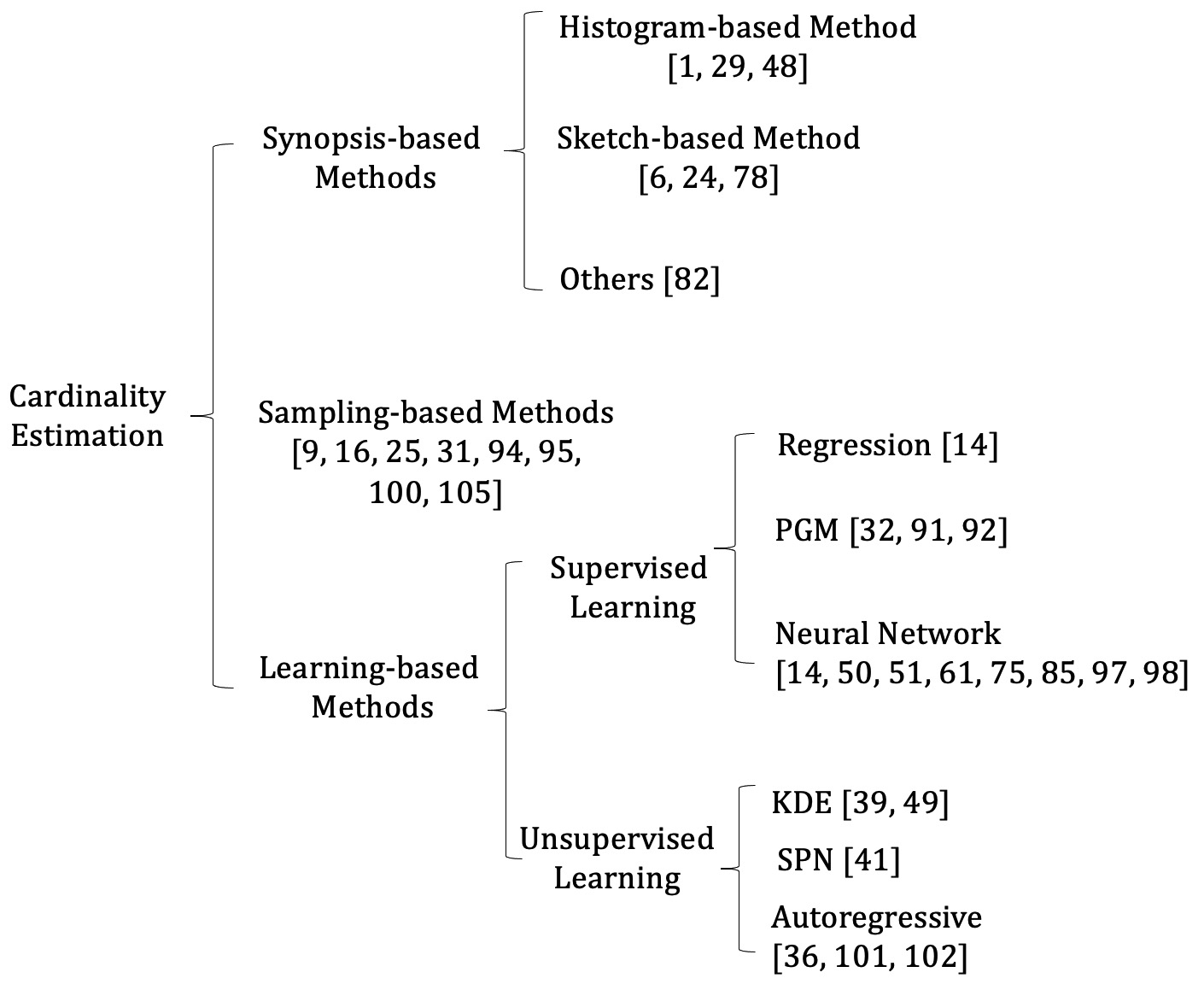}
    \caption{A classification of cardinality estimation methods}
    \Description{Cardinality estimation methods}
    \label{classification_ce}
  \end{figure}

At present, there are three major strategies for cardinality estimation as shown in Figure \ref{classification_ce}.
We only list some representative work for each category. 
Every method tries to approximate the distribution of data well with less storage. 
Some proposed methods combine different techniques, e.g.,~\cite{CE_L_PGM_TzoumasDJ11,CE_L_PGM_TzoumasDJ13}.
\subsection{Synopsis-based methods}\label{Synopsis_CE}
Synopsis-based methods introduce new data structures to record the statistics information. Histogram and sketch are the widely adopted forms.
A survey on synopses has been proposed in 2012~\cite{CE_HS_SV_CormodeGHJ12}, which focuses on distinguishing aspects of synopses that are pertinent to Approximate Query Processing (AQP). 
\subsubsection{Histogram}  
There are two histogram types: \textit{1}-dimensional and \textit{d}-dimensional histograms, 
where \(d \ge 2\). \textit{d}-dimensional histograms can capture the correlation between different attributes.

A \textit{1-dimensional histogram} on attribute \textit{a} is constructed by partitioning the sorted tuples into \(B\) (\(\ge1\)) mutually disjoint subsets, called \textit{buckets}
and approximates the frequencies and values in each bucket in some common fashion, e.g., uniform distribution and continuous values.
A \textit{d-dimensional histogram} on an attribute group \textit{A} is constructed by partitioning the joint data
distribution of \textit{A}. Because there is no order between different attributes, the partition rule needs to be more intricate.
In 2003, \citet{CE_HS_SV_Ioannidis03} present a comprehensive survey on histograms following the classification method in~\cite{HS_PoosalaIHS96}.
\citet{HS_SV_ML_GunopulosKTD05} also propose a survey in 2003, which focuses on the work used to estimate the selectivity over multiple attributes. 
They summarize the multi-dimensional histograms and kernel density estimators.
After 2003, the work in histograms can be divided into three categories: 
(1) fast algorithm for histogram construction~\cite{HS_FS_GuhaKS06,HS_FS_HalimKY09,HS_FS_HalimKY10,HS_CS_IndykLR12,HS_CS_AcharyaDHLS15}; 
(2) new partition methods to divide the data into different buckets to achieve better accuracy~\cite{HS_EavisL07,HS_ToCS13,HS_LinZPS15}; 
(3) histogram construction based on query feedback~\cite{FDBK_LimWV03,FDBK_SrivastavaHMKT06,FDBK_KaushikS09}.
Query feedback methods are also summarized in~\cite{CE_HS_SV_CormodeGHJ12} (Section 3.5.1.2) and readers can refer to it for details.

\citet{HS_FS_GuhaKS06} analyze the previous algorithm, VODP~\cite{VODP_JagadishKMPSS98} and, find some calculations on the minimal sum-of-squared-errors (SSE) can be reduced.
They design an efficient algorithm AHistL-\(\Delta\) with time complexity \(O(n+B^3(\lg n+\epsilon^{-2}))\) while VODP takes \(O(n^2B)\), where \(n\) is the domain size, \(B\) is the number of buckets, and \(\epsilon\) is a precision parameter.
\citet{HS_FS_HalimKY09,HS_FS_HalimKY10} propose GDY, a fast histogram construction algorithm based on greedy local search. 
GDY generates good sample boundaries, which then are used to construct \(B\) final partitions 
optimally using VODP. This study compares GDY variants with AHistL-\(\Delta\)~\cite{HS_FS_GuhaKS06} 
in minimizing the total errors of all the buckets and shows its superiority in resolving the efficiency-quality trade-off.
Instead of scanning the whole dataset~\cite{HS_FS_GuhaKS06}, \citet{HS_CS_IndykLR12} design a greedy algorithm to construct 
the histogram on the random samples from dataset with time complexity \(O((B^5/\epsilon^8)\log^2n)\) and sample complexity \(O((B/\epsilon)^2\log n)\).
\citet{HS_CS_AcharyaDHLS15} study the same problem with~\cite{HS_CS_IndykLR12} and propose a merging algorithm with time complexity \(O(1/\epsilon^2)\). Methods in~\cite{HS_FS_GuhaKS06,HS_CS_IndykLR12,HS_CS_AcharyaDHLS15} can be
extended to approximate distributions by piecewise polynomials.

Considering the tree-based indexes divide the data into different segments (nodes), which is quite similar with buckets 
in the histogram, \citet{HS_EavisL07} build the multi-dimensional histogram based on R-tree. They first build a native 
R-tree histogram on the Hilbert sort of data, and then, propose a sliding window algorithm to enhance the naive histogram 
under a new proposed metric, which seeks to minimize the dead space between bucket points. 
\citet{HS_LinZPS15} design a two-level histogram for one attribute, which is 
quite similar to the idea of the B-tree index. The first level is used to locate which leaf histograms to be used, 
and the leaf histograms store the statistics information.
\citet{HS_ToCS13} construct a histogram based on the principle of minimizing the entropy reduction of the histogram.  
They design two different histograms for the equality queries and an incremental algorithm to construct the histogram. 
However, it only considers the one-dimensional histogram and does not handle range queries well.

\subsubsection{Sketch}
Sketch models a column as a vector or matrix to calculate the distinct count (e.g., HyperLogLog~\cite{HyperLogLog}) or frequency of tuples (e.g., Count Min~\cite{CountMin})
on a value. \citet{Sketch_RusuD08} summarize how to use different sketches to estimate the join size.
This work considers the case of two tables (or data streams) without filters.
The basic idea of them is: (1) building the sketch (a vector or matrix) on the join attribute, while ignoring all the
other attributes, (2) estimating the join size based on the multiplication of the vectors or matrices.
These methods only support the equi-join and join on single column.
As shown in~\cite{CE_SA_VengerovMZC15}, a possible method introducing one filter in sketch is to build an imaginary table which 
only consists of the join value of tuples which satisfy the filter. However, this makes the estimation drastically worse.
Skimmed sketch~\cite{GangulyGR04} is based on the idea of bifocal sampling~\cite{CE_SA_GangulyGMS96} to estimate the join size.
However, it requires knowing frequencies of the most frequent join attribute values. Recent work~\cite{Sketch_CaiBS19} on join size estimation introduces the sketch
to record the degree of a value. 

\subsubsection{Other Techniques}
TuG~\cite{SY_GR_SpiegelP06} is a graph-based synopsis. The node of TuG represents a set of tuples from the same table or a set of values for the same attribute. 
The edge represents the join relationship between different tables or between attributes and values. 
The authors adopt a three-step algorithm to construct TuG and introduce the histogram to summarize the value distribution in a node.  
When a new query comes, the selectivity is estimated by traversing TuG. 
The construction process is quite time-consuming and cannot be used in a large dataset. Without the relationship between different tables, TuG cannot be built.
\subsection{Sampling-based Methods}\label{sampling}
Synopsis-based methods are quite difficult to capture the correlation between different tables. 
Some researchers try to use a specific sampling strategy to collect a set of 
samples (tuples) from tables, and then run the (sub)query over samples to estimate the cardinality.
As long as the distribution of the obtained samples is close to the original data, the cardinality estimation is believable.
Thus, lots of work have been proposed to design a good sampling approach, from independent sampling to correlated sampling technique.
Sampling-based methods also are summarized in~\cite{CE_HS_SV_CormodeGHJ12}. After 2011, there are numerous studies that utilize the sampling techniques. Different with~\cite{CE_HS_SV_CormodeGHJ12},
 we mainly summarize the new work. Moreover, we review the work according to their publishing time and present the relationship between them, i.e., which shortages of the previous work 
 the later work tries to overcome.

In 1993, \citet{CE_SA_HaasNSS93} analyze the six different fixed-step (a pre-defined sample size) sampling methods for the equi-join queries. 
They conclude that if there are some indexes built on join keys, page-level sampling combining the index is the best way. Otherwise, the page-level cross-product sampling is the most efficient way.
Then, the authors extend the fixed-step methods to fixed-precision procedures.

\citet{CE_SA_GangulyGMS96} introduce bifocal sampling to estimate the size of an equi-join. They classify values of the join attribute in each relation into two groups, sparse (\textit{s}) and dense (\textit{d}) based on their frequencies. 
Thus, the join type between tuples can be \textit{s-s}, \textit{s-d}, \textit{d-s}, and \textit{d-d}. 
The authors first adopt \textit{t\_cross} sampling~\cite{CE_SA_HaasNSS93} to estimate the join size of \textit{d-d}
, then adopt \textit{t\_index} to estimate the join size of the remaining cases, and finally add all the estimation as the join size estimation. However, it needs an extra pass to determine the frequencies of different values and needs
indexes to estimate the join size for \textit{s-s}, \textit{s-d}, and \textit{d-s}. Without indexes, the process is time-consuming.

End-biased sampling~\cite{CE_SA_EstanN06} stores the \((v,f_v)\) if \(f_v \ge T\), where \(v\) is a value in the join attribute domain, \(f_v\) is the number of tuples with value \(v\), and \(T\) is a defined threshold.
It applies a hash function \(h(v):v \mapsto [0,1]\). If \(h(v) \le \frac{f_v}{T}\), it stores \((v,f_v)\) or not. 
Different tables adopt the same hash function to correlate their sampling decisions for tuples with low frequencies. Then the join size can be estimated using stored \((v,f_v)\) pairs. 
However, it only supports equi-join on two tables and cannot handle other filter conditions.
Notice, end-bias sampling is quite similar to bifocal sampling. 
The difference is: the former uses a hash function to sample correlated tuples and the latter uses the indexes. Both of them require an extra pass through the data to compute the frequencies of the join attribute values.

\citet{CE_SA_YuHLCZ13} introduce correlated sampling as a part of CS2 algorithm. They (1) choose one of the tables in a join graph as the source table \(R_1\), (2) use a random sampling method to obtain sample set \(S_1\)
for \(R_1\) (mark \(R_1\) as visited), (3) follow an unvisited edge \(<R_i, R_j>\) (\(R_i\) is visited) in the join graph and collect the tuples from \(R_j\) which are joinable with tuples in \(S_i\) as \(S_j\), and 
(4) estimate the join size over the samples. To support the query without source tables, they propose a reverse estimator, which tracks to the source tables to estimate the join size. However, due to the walking through the join graph
many times, it is time-consuming without indexes. Furthermore, it requires an unpredictable large space to store the samples. 

\citet{CE_SA_VengerovMZC15} propose a correlated sampling method without the prior knowledge of frequencies of join attributes, like in~\cite{CE_SA_GangulyGMS96, CE_SA_EstanN06}. 
A tuple with join value \(v\) is included  in the sample set if \(h(v) < p\), where \(p = \frac{n}{T}\), \(h(v)\) is a hash function similar in~\cite{CE_SA_EstanN06}, 
\(n\) is the sample size, and \(T\) is the table size. Then, we can use obtained samples to estimate the join size and handle specified filter conditions. 
Furthermore, the authors extend the method into more tables join and complex join conditions.
In most cases, the correlated sampling has lower variance than independent Bernouli sampling (\textit{t\_cross}), 
but when the values of many join attributes occur with large frequencies, the Bernouli sampling is better. One possible solution the authors propose is to adopt a one-pass algorithm to detect the values with high frequencies,which is back to
the method in~\cite{CE_SA_EstanN06}.

Through experiments, \citet{CE_SA_ChenY17} conclude that there does not exist one sampling method suitable for all cases. 
They propose a two-level sampling method, which is based on the independent Bernouli sampling, end-bias sampling~\cite{CE_SA_EstanN06}, and correlated sampling~\cite{CE_SA_VengerovMZC15}.
Level-one sampling samples a value \(v\) from join attribute domain into value set (\(V\)), if \(h(v) < p_v\). \(h\) is a hash function similar to~\cite{CE_SA_EstanN06}, \(p_v\) is a defined probability for value \(v\).
Before level-two sampling, they sample a random tuple, called the sentry, for every \(v\) in \(V\) into tuple set. Level-two sampling samples tuples with value \(v\) (\(v \in V\)) with probability \(q\). 
Then, we can estimate the join size by using the tuple samples. Obviously, the first level is a correlated sampling and the second level is independent Bernouli sampling.
The authors analyze how to set the \(p_v\) and \(q\) according to different join types and the frequencies of values in join attributes.

\citet{CE_SA_WangC20} extend~\cite{CE_SA_ChenY17} to a more general framework in terms of five parameters.
Based on the new framework, they propose a new class of correlated sampling methods, called CSDL, which is based on 
the discrete learning algorithm. A variant of CSDL, CSDL-Opt has outperformed~\cite{CE_SA_ChenY17} when the samples are small or
join value density is small. 

\citet{Sampling_WuNS16} adopt the online sampling to correct the possible errors in the plan generated by the optimizers. 

\subsection{Learning-based Methods}
Due to the capability of the learning-based methods, many researchers have introduced a learning-based model to capture the distribution and correlations of data. 
We classify them into: (1) supervised methods, (2) unsupervised methods.

\begin{table*}[t]
  \setlength{\tabcolsep}{1mm}{
    \renewcommand\arraystretch{0.8}
    \newcommand{\tabincell}[2]{\begin{tabular}{@{}#1@{}}#2\end{tabular}}
    \caption{Learning-based methods for cardinality estimation.}
    \label{tab:L_CE}
    \begin{tabular}{cccccccccc}
      \toprule
      Refs & Model & Model Count & Encoding & Multi-columns & Multi-tables & UDF & Workload Shift\\
      \midrule
      \cite{CE_L_SP_MalikBC07} & LR & 1 Model/1 Template & predicates, arguments & \checkmark &  \checkmark & \checkmark & $\times$  \\
      \cite{CE_L_ParkZM20} & MixModel & 1 Model & predicates & \checkmark &  $\times$ & $\times$ & \checkmark   \\
      \cite{CE_L_PGM_TzoumasDJ11,CE_L_PGM_TzoumasDJ13} & BN & 1 Model & predicates & \checkmark &  \checkmark & $\times$ & $\times$   \\
      \cite{CE_L_PGM_HalfordSM19} & BN & 1 Model/1 Table & predicates & \checkmark &  $\times$ & $\times$ & $\times$   \\
      \cite{CE_Learning_LakshmiZ98} & NN & 1 Model/1 UDF & arguments & $\times$ &  $\times$ & \checkmark & $\times$   \\
      \cite{CE_L_SP_LiuXYCZ15} & NN & 1 Model & predicates & \checkmark &  $\times$ & $\times$ & $\times$   \\
      \cite{CE_L_SP_WuJAPLQR18} & NN/PR/MLR & 1 Model/1 Subquery & predicates, input cardinalities & \checkmark &  \checkmark & \checkmark & $\times$  \\
      \cite{CE_L_SP_KipfKRLBK19} & MSCN & 1 Model & predicates, tables, joins  & \checkmark &  \checkmark & $\times$ & $\times$  \\
      \cite{CE_L_SP_DuttWNKNC19} & Tree-Ensemble/NN & 1 Model & predicates & \checkmark & $\times$ & $\times$ & $\times$   \\
      \cite{CE_L_SP_Lucas20,CE_L_SP_WoltmannHTHL19} & NN & 1 Model/1 Template & predicates & \checkmark &  \checkmark & $\times$ & $\times$   \\
      \cite{CE_L_SP_Ortiz19} & DNN/RNN/Tree & 1 Model & predicates, tables, joins & \checkmark &  \checkmark & $\times$ & $\times$   \\
      \cite{CE_L_SP_SunL20} & tree-LSTM & 1 Model & predicate, operator, metadata & \checkmark &  \checkmark & $\times$ & $\times$   \\
      \midrule
      \cite{CE_L_KDE_HeimelKM15} & KDE & 1 Model & samples & \checkmark &  $\times$ & $\times$ & \checkmark  \\
      \cite{CE_L_KDE_KieferHBM17} & KDE & \tabincell{c}{1 Model \\ 1 Model / 1 Table} & samples & \checkmark &  \checkmark & $\times$ & \checkmark   \\
      \cite{CE_L_US_HilprechtSKMKB20} & SPN & 1 Model & tuples; predicates & \checkmark &  \checkmark & $\times$ & \checkmark   \\
      \cite{CE_L_US_YangLKWDCAHKS19,CE_L_US_HasanTAK020} & Autoregression & 1 Model & tuples; predicates & \checkmark &  $\times$  & $\times$ & \checkmark  \\
      \cite{CS_L_U_abs-2006-08109} & Autoregression & 1 Model & tuples; predicates & \checkmark &  \checkmark & $\times$ & \checkmark  \\
      \bottomrule
    \end{tabular}}
  \end{table*}

\subsubsection{Supervised Methods} \label{SM}
\begin{table*}[t]
  \setlength{\tabcolsep}{1mm}{
    \renewcommand\arraystretch{0.8}
    \newcommand{\tabincell}[2]{\begin{tabular}{@{}#1@{}}#2\end{tabular}}
    \caption{A preliminary comparison in different methods for cardinality estimation.}
    \label{tab:CE_C}
    \begin{tabular}{cccccccc}
      \toprule
      Methods & Workload Shift & Data Change & Update Time & Storage Usage & Multi-Columns & Multi-Tables \\
      \midrule
      \textit{1}-histogram\cite{CE_HS_SV_Ioannidis03} & $\times$ & $\times$ & Short & Small & $\times$ & $\times$  \\
      \textit{d}-histogram~\cite{HS_SV_ML_GunopulosKTD05} & $\times$ & $\times$ & Short & Large & \checkmark & $\times$  \\
      Sampling~\cite{Sampling_WuNS16,CE_SA_ChenY17,CE_SA_WangC20} & \checkmark & $\times$ & Medium & Large & \checkmark & \checkmark  \\
      Supervised Learning~\cite{CE_L_SP_KipfKRLBK19,CE_L_SP_SunL20} & $\times$ & $\times$ & Long & Small & \checkmark & \checkmark  \\
      Unsupervised Learning~\cite{CS_L_U_abs-2006-08109,CE_L_US_HilprechtSKMKB20} & \checkmark & $\times$ & Long & Small & \checkmark & \checkmark \\
      \bottomrule
    \end{tabular}}
  \end{table*}
\citet{CE_L_SP_MalikBC07} group queries into templates, and adopt machine learning
techniques (e.g., linear regression model, tree models) to learn the distribution of query result sizes for each
family. The features used in it include query attributes, constants, operators, aggregates, and arguments to UDFs.

\citet{CE_L_ParkZM20} propose a model, QuickSel, in query-driven paradigm, which is similar to~\cite{FDBK_LimWV03,FDBK_SrivastavaHMKT06,FDBK_KaushikS09}, 
to estimate the selectivity of one query. Instead of adopting the histograms, QuickSel introduces the uniform mixture models to represent the distribution of the data.
They train the model by minimizing the mean squared error between the mixture model and a uniform distribution.

\citet{CE_L_PGM_TzoumasDJ11,CE_L_PGM_TzoumasDJ13} build 
a Bayesian network and decompose the complex statistics over multiple attributes into small one-/two-dimensional statistics, 
which means the model captures dependencies between two relations at most.
They build the histograms for these small dimensional statistics and adopt a dynamic programming to calculate the selectivity for the new queries.
Different with previous method~\cite{CE_L_PGM_GetoorTK01}, it can handle more general joins and has a more efficient construction algorithm because of capturing smaller dependencies. 
However, the authors do not verify their method with multiple tables join and in large dataset. Moreover, constructing the two-dimensional statistics with attributes from different tables needs the join operation.
\citet{CE_L_PGM_HalfordSM19} also introduce a method based on Bayesian network. To construct the model quickly, they only factorize the distribution of attributes
inside each relation and use the previous assumptions for joins. 
However, they do not present how well their method compared with \cite{CE_L_PGM_TzoumasDJ11,CE_L_PGM_TzoumasDJ13}.

In 1998, \citet{CE_Learning_LakshmiZ98} first introduce the NN into 
the cardinality estimation of user defined function (UDF), which the histograms and other statistics cannot support. They design a two-layer neural network (NN) and employ the back propagation
to update the model. 

\citet{CE_L_SP_LiuXYCZ15} formalize a selectivity function, \(Sel: R^{2N} \mapsto R, (l_1,u_1,...,l_n,u_n) \mapsto c\), where \(N\) is the number of attributes, \(l_i\) and \(u_i\)
is the lower and upper bound on \(i^{th}\) attribute for a query. They employ a 3-layer NN to learn the selectivity function. To support \(>\) and \(<\), they add \(2N\) small
NNs to produce \(l_i\) and \(u_i\). 

\citet{CE_L_SP_WuJAPLQR18} use a learning-based method for workload in shared clouds, where the queries are often recurring and overlapping in nature. 
They first extract overlapping sub-graph templates in multiple query graphs. Then, they learn the cardinality models for those sub-graph templates.

\citet{CE_L_SP_KipfKRLBK19} introduce the multi-set convolutional network (MSCN) model to estimate the cardinality of correlated joins. 
They represent a query as a collection of a set of tables \(T\), joins \(J\), and predicts \(P\) and build the separate 2-layer NN for each of them.
Then, the outputs of three NNs are concatenated after the averaging operation and fed into the final output network. 
Deep sketch~\cite{CE_L_SP_Kipf19} is built on~\cite{CE_L_SP_KipfKRLBK19} and is a wrapper of it.

\citet{CE_L_SP_DuttWNKNC19} formalize the estimation as a function similar to~\cite{CE_L_SP_LiuXYCZ15}, and they consider it as a regression problem. 
They adopt two different approaches for the regression problem, NN-based methods and tree-based ensembles. Different with~\cite{CE_L_SP_LiuXYCZ15}, 
the authors also use histograms and domain knowledge (e.g., AVI, EBO, and MinSel) as the extra features in the models, which improve the estimation accuracy.
Due to the domain knowledge quickly updated when the data distribution changes, the model is robust to the updates on the datasets. 

\citet{CE_L_SP_WoltmannHTHL19} think building a single NN, called global model, over the entire database schema has the sparse encoding and needs 
numerous samples to train the model. Thus, they build different models, called local models, for different query templates.
Every local model adopts multi-layer perceptrons (MLP) to produce the cardinality estimation. To collect the true cardinality, many sample queries are 
issued during the training process, which is time-consuming. Furthermore, \citet{CE_L_SP_Lucas20} introduce the method of pre-aggregating the base data using the data cube concept 
and execute the example queries over this pre-aggregated data. 

\citet{CE_L_SP_Ortiz19} empirically analyze various of deep learning approaches used in cardinality estimation, including deep neural network (DNN) and recurrent neural network (RNN).
The DNN model is similar with~\cite{CE_L_SP_WoltmannHTHL19}. To adopt RNN model, the authors focus on left-deep plans and model a query as a series of 
actions. Every action represents an operation (i.e., selection or join). In each timestamp \(t\), the model receives two inputs: \(x_t\), the encoding of \(t^{th}\) of operation, and 
\(h_{t-1}\), the generated hidden state from timestamp \(t-1\), which can be regarded as the encoding of a subquery and captures the important details about the intermediate results.

\citet{CE_L_SP_SunL20} introduce a tree-LSTM model to learn a representation of an operator and add an estimation layer upon the tree-LSTM model 
to estimate the cardinality and cost simultaneously. 

\subsubsection{Unsupervised Methods} \citet{CE_L_KDE_HeimelKM15} introduce the Kernel Density Estimator (KDE) into estimating the selectivity on single table with multiple predicates.
They first adopt the Gaussian Kernel and the bandwidth obtained by a certain rule to construct the initial KDE, and then they use the history queries 
to choose the optimal bandwidth by minimize the estimation error using initial KDE. To support the shifts in workload and dataset, they update the bandwidth after each incoming query and  design the new sample maintenance method for insert-only workload and updates/deletions workload. 
Furthermore, in 2017, \citet{CE_L_KDE_KieferHBM17} extend the method into estimating the selectivity of join. They design two different models: single model over the 
join samples and the models over the base tables, which does not need the join operation and estimates the selectivity of join with 
the independent assumption.

\citet{CE_L_US_YangLKWDCAHKS19} propose a model called Naru, which adopts the deep autoregressive model to produce \(n\) conditional densities \(\hat{P}(x_i|x_{<i})\) on a set of \(n\)-dimensional tuples.
Then they estimate the selectivity using the product rule: 
\begin{equation}
    \begin{aligned}
    \hat{P}(\textbf{x})=&\hat{P}(x_1,x_2,...,x_n)\\
    =&\hat{P}(x_n|x_1,...,x_{n-1})\hat{P}(x_{n-1}|x_1,...,x_{n-2})...\hat{P}(x_2|x_1)\hat{P}(x_1)
    \end{aligned}
\end{equation}
To support range conditions, they introduce a progressive sampling method by sampling points
from more meaningful region according to the trained autoregressive model, which is robust to the skewed data.
Furthermore, they adopt the wildcard-skipping to handle wildcard condition.

\citet{CE_L_US_HasanTAK020} also adopt the deep autoregressive models and introduce an adaptive sampling method to support range queries.
Compared with the Naru, the authors adopt the binary encoding method and the sampling process runs
parallelly, which leads the model is smaller than Naru and makes the inference faster. Besides, it can incorporate with the workload
by assigning the tuples with weights according to the workload when defining the cross-entropy loss function. 

\citet{CE_L_US_HilprechtSKMKB20} introduce the Relational Sum Product Network (RSPN) to capture
the distribution of single attributes and the joint probability distribution. They focus on Tree-SPNs, where one leaf
is the approximation of a single attribute, the internal node is Sum node (splitting the rows into clusters) or 
Product node (splitting the columns of one cluster). 
To support cardinality estimation of join, they build the RSPN over the join results.

\citet{CS_L_U_abs-2006-08109} extend their previous work, Naru, to support joins. 
They build an autoregressive model over the full outer join of all tables.
They introduce the lossless column factorization for large-cardinality columns and employ
the join count table to support any queries on the subset of tables.
\subsection{Our Insights}
\subsubsection{Summaries}
The basic histogram types (e.g., equi-width, equi-depth, \textit{d}-dimensional) 
have been introduced before 2000. Recent studies mainly focus on how to quickly construct the histograms and to improve the accuracy of them.
Updating the histograms by query feedback is a good approach to improve the quality of histograms. However, there are still two limitations in the histograms: 
(1) the storage size increases dramatically when building a \textit{d}-dimensional histograms; (2) histograms cannot capture the correlation between attributes from different tables.
If building a histogram for the attributes from different tables, the join operation is required, like in~\cite{CE_L_PGM_TzoumasDJ11,CE_L_PGM_TzoumasDJ13} and it is difficult to update
this histogram. 
Sketch can be used to estimate the distinct count of an attribute or the cardinality of equi-join results. 
However, it cannot support more general cases well, e.g., join with filters. Synopsis-based methods cannot estimate the size of
final or intermediate relations when one or both of the child
relations is an intermediate relation.

Sampling is a good approach to capture the correlations between different tables.
However, when the tuples in tables have been updated, the samples may become out-of-date.
Sampling-based methods also suffer from the storage used to store the samples and the time used to retrieve the samples, 
especially when the original data is numerous. Furthermore, current sampling methods only support the equi-join.

The supervised learning methods are mostly query-driven, which means the model is trained for a specific workload.
 If the workload shifts, the model needs to be retrained.
Thus, the data-driven (unsupervised learning) approaches come out, which still can estimate the cardinality even if the workload shifts.
As shown in~\cite{CE_L_US_YangLKWDCAHKS19} (Section 6.3), Naru is robust to workload shift while MSCN and KDE are sensitive to the training queries.
Moreover, both of the supervised and unsupervised learning methods suffer from the data change. As presented in~\cite{CS_L_U_abs-2006-08109} (Section 7.6) and 
~\cite{CE_L_SP_SunL20} (Section 7.5), both of them are sensitive to data change and the models will be updated in an incremental mode or retrained from scratch. However, 
they only consider that new tuples are appended into one table and there does not exist delete or update operation.

Due to the difference in the experiment settings, we only present a preliminary comparison between the methods for cardinality estimation as shown in Table \ref{tab:CE_C}.
Sometimes, the size of learning-based methods is still not small as presented in~\cite{CS_L_U_abs-2006-08109}. 
The state-of-the-art method is \cite{CS_L_U_abs-2006-08109}, which shows superiority compared with other methods in their experiments.

\subsubsection{Possible Future Directions}\label{PFD_CE} There are several possible directions as follows: 
\begin{enumerate} 
  \item \textbf{Learning-based methods.} Many studies on cardinality estimation are learning-based methods in last two years. The learning-based models currently integrated a real system 
  are the light model or one model for one (sub-)query graph~\cite{CE_L_SP_WuJAPLQR18,CE_L_SP_DuttWNKNC19}, which can be trained and updated quickly. 
  However, the accuracy and generality of these models are limited. More complex models (achieve a better accuracy) still suffer from the long training time and update time.
  Training time is influenced by the hardware. 
  For example, it only takes several minutes in~\cite{CS_L_U_abs-2006-08109}, while it is 13 hours in~\cite{CE_L_SP_SunL20} using a GPU with relatively poor performance.
  A database instance, especially in the cloud environment, is in a resource-constrained environment. How to train the model efficiently should be considered.
  The interaction between the models and the optimizer also needs to be considered, which should not be with too much overhead on the database systems~\cite{CE_L_SP_DuttWNKNC19}.
  As presented above, current proposed methods for data change cannot handle delete or update operation. A possible method is to adopt the idea of active learning to update the model~\cite{ActiveL_MaDDS20}.
  \item \textbf{Hybrid methods.} Query-driven methods are sensitive to workload. Although data-driven methods can support more queries, it may not achieve the best accuracy for all queries.
  How to combine two methods in these two different catalogs is a possible direction. Actually, the previous query-feedback histograms is an instance of this case. 
  Another interesting thing is that utilizing the query feedback information will help the model be aware of the data change.
  \item \textbf{Experimental Study.} Although many methods have been proposed, it lacks of experimental studies to verify these methods. Different methods have different characteristics 
  as shown in Table \ref{tab:CE_C}. It is crucial to conduct a comprehensive experimental study for proposed methods.
  We think the following aspects should be included: (1) is it easy to integrate the method into a real database system; 
  (2) what is the performance of the method under different workload patterns (e.g., static or dynamic workload, OLTP or OLAP) and different data scales and distributions;
  (3) researchers should pay attention to the trade-off between storage usage and accuracy of candidate estimation and the trade-off between efficiency of model update and the accuracy.
\end{enumerate}

\section{Cost Model}\label{CM}
\begin{figure}[t]
    \centering
    \includegraphics[width=\linewidth]{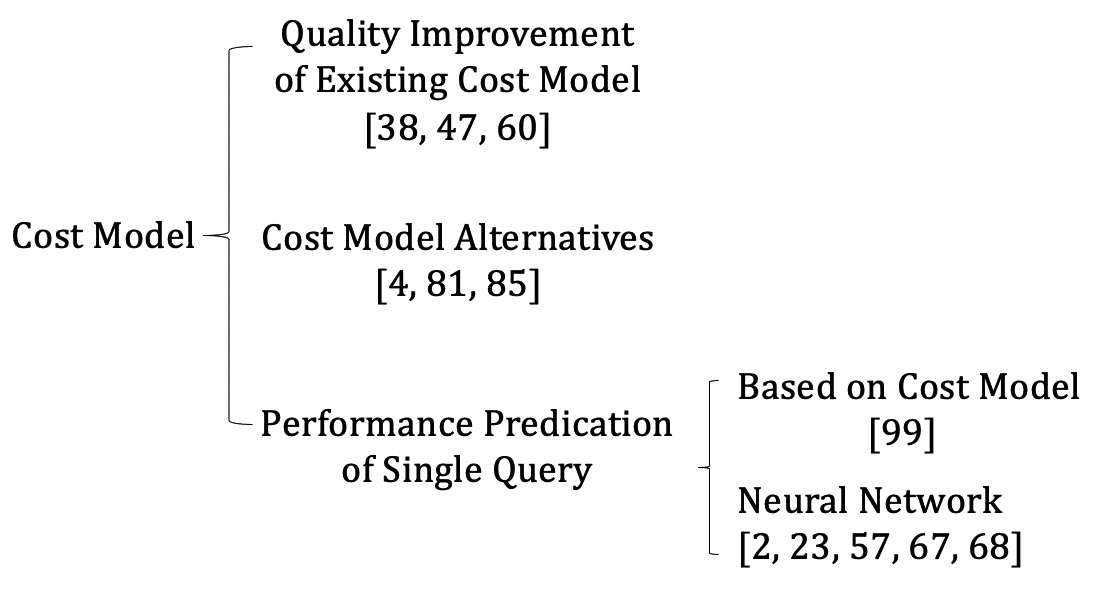}
    \caption{A classification of cost estimation methods}
    \Description{cost estimation methods}
    \label{classification_cm}
  \end{figure}
In this section, we present the researches proposed to solve the limitations in the cost model. We classify the methods into three groups: (1) improving the capability of the existed cost model, (2) building a new cost model, and (3) predicting the query performance.
We include the work on the single query performance prediction. Because the cost used in the optimizer is the metric for performance. 
These methods are possibly integrated into the cost-based optimizer and replace the cost model
to estimation the cost of a (sub)plan, like in~\cite{CM_L_MarcusNMZAKPT19}. 
However, we do not consider the query performance prediction under concurrent context (e.g., ~\cite{PF_C_ZhouSLF20}). On the one hand, the concurrent queries
existing during the optimization may be quite different with queries during the execution process. On the other hand, it also needs to collect more information than the model predicting the performance of a
single query. We list the representative work of in the cost model in Figure \ref{classification_cm}.
\subsection{Quality Improvement of Existing Cost Model}
Several studies try to estimate the cost of UDF~\cite{CM_T_BoulosO99, CM_HeLS04, CM_L_HeLS05}. \citet{CM_T_BoulosO99} execute the UDF several times with 
different input values, collect the different costs, and then use these costs to build a multi-dimensional histogram. The histogram is stored in a tree
structure. When estimating the UDF with specific parameters, traverse the tree top-down to get the estimated cost to locate the leaf with similar parameters with inputs.
However, this method needs to know the upper and lower bounds of every parameter and it cannot solve the complex relation between input parameters
and the costs. Unlike the static histogram used in~\cite{CM_T_BoulosO99}, \citet{CM_HeLS04} introduce a dynamic quadtree-based approach to store the UDF execution information.
When a query is executed, the actual cost of executing the UDF is used to update the cost model. \citet{ CM_L_HeLS05} introduce a memory-limited K-nearest neighbors (MLKNN) method.
They design a data structure, called PData, to store the execution cost and 
a multidimensional index used for fast retrieval k nearest PData for a given query point (parameter in UDF) and fast insertion of new PData.

\citet{CM_LiuB15} introduce a cost model for hash-based join for main-memory database.
They model the response time of a query as being proportional to the number of operations weighted by the costs of four basic access patterns.
They first adopt the microbenchmarks to get the cost of each access pattern and then model the cost of sequential scan, hash join, hash join with different orders
by the basic access patterns.

Most of the previous cost models only consider the execution cost, which may be not reasonable in the cloud environments. The users of the cloud database systems care about the economic cost.
\citet{CM_L_KarampaglisGM14} first propose a bi-objective query cost model, which is used to derive running time and monetary cost together in the multi-cloud environment.
They model the execution time based on the method in~\cite{CM_L_WuCZTHN13}. For economic cost estimation, they first model the charging policies and estimate the monetary cost by
combining the policy and time estimation.

\subsection{Cost Model Alternatives}
The cost model is a function mapping the (sub)plan with annotated information to a scalar (cost). 
Because a neural network on data primarily approximates the unknown underlying mapping function from inputs to outputs, 
most of the methods used to replace the origin cost model are learning-based, especially NN-based. 

\citet{CM_T_boulos1997neural} firstly introduce the neural network for cost evaluation. They design two different models: a single large neural network for
every query type and a single small neural network for every operator. In the first model, they also train another model to classify a query in a certain type. The output
of the first model is the cost of a (sub)plan, while the second model needs to add up the outputs from small models to get the cost.

\citet{CE_L_SP_SunL20} adopt a tree-LSTM model to learn the presentation of an operator and add an estimation layer upon the tree-LSTM model to 
estimate the cost of the query plan.

Due to the difficulty in collecting statistics and the needs of picking the resources in big data systems, particularly in modern cloud data services,
\citet{CM_L_SiddiquiJQPL20} propose a learning-based cost model and integrate it into the optimizer of SCOPE~\cite{SCOPE_ChaikenJLRSWZ08}. 
They build large number of small models to predict the costs of common (sub)queries, which are extracted from the workload history. The features encoded
into the models are quite similar with~\cite{CE_L_SP_WuJAPLQR18}. Moreover, to support resource-aware query planning, they add number of partitions allocated to the operator into the features.
In order to improve the coverage of the models, they introduce operator-input models and operator-subgraphApprox
models and employ a meta-ensemble model to combine the models above as the final model.

\subsection{Query Performance Prediction} The performance of the one query mainly refers to the latency.
\citet{CM_L_WuCZTHN13} adopt an offline profiling to calibrate the coefficients in the cost model under a specific hardware and software conditions. Then, they adopt the 
sampling method to obtain the true cardinalities of the physical operators to predict the execution times.

\citet{CM_P_GanapathiKDWFJP09} adopt the Kernel Canonical Correlation Analysis (KCCA) into the resource estimation, e.g. CPU time. 
They only model the plan level information, e.g. the number of each physical operator type and their cardinality, which is too vulnerable.

To estimate the resources (CPU time and logical I/O times), \citet{CM_L_P_LiKNC12} train a boosted regression tree for every operator 
in the database and the consumption of the plan is the sum of the operators’. To make the model more robust, they train 
a separate scaling function for every operator and combine scaling functions with the regression models to handle the cases when 
the data distribution, size or queries’ parameters are quite different with the training data. Different with~\cite{CM_P_GanapathiKDWFJP09}, this is an operator-level model. 

\citet{CM_P_LAkdereCRUZ12} propose the learning-based models to predict the query performance. They first design a plan-level model if the workload is known in advance and an operator-level model.
Considering the plan-level model makes highly accurate prediction and the operator-level generalizes well, for queries with low operator-level prediction
accuracy, they train models for specific query subplans using plan-level modeling and compose both types of models to predict the performance of the entire plan.
However, the models adopted are linear.

\citet{CM_P_MarcusP19} introduce a plan-structure neural network to predict the query's latency. They design a small neural network, called neural unit, for every logic operator and any instance of the same logic operator
shares the same network. Then, these neural units are combined into a tree shape according to the plan tree. The output of one neural unit consists of two parts, the latency of current operator and the information sent
to its parent node. The latency of the root neural unit of a plan is the plan's latency.

Neo~\cite{CM_L_MarcusNMZAKPT19} is a learning-based query optimizer, which introduces a neural network, called value network, to estimate the latency of (sub)plan.
The query-level encoding (join graph and columns with predicts) is fed through several full-connected layers and then concatenated with the plan level encoding, which is a tree vector to represent the
physical plan. Next, the concatenated vector is fed into a Tree Convolution and another several full-connected layers to predict the latency of the input physical plan.

\subsection{Our Insights}
\subsubsection{Summaries} The methods trying to improve the existing cost model focus on different aspects, e.g., UDFs, hash join in main memory. 
These studies leave us an important lesson: when introducing a new logical or physical operator, 
or re-implementing the existing physical operators, we should consider how to add them into 
the optimization process and design the corresponding cost estimation formulas 
for them (e.g.,~\cite{COST_LeekaR19,JOIN_T_TD_NamH020}).

Leaning-based methods adopt the model to capture the complex relationship between cost and the factors 
while the traditional cost model is defined as a certain formula by the database experts.
The NN-based methods used to predict the performance, estimate cost, and estimate cardinality in Section \ref{SM} 
are quite similar in the features and models selection. For example, \citet{CE_L_SP_SunL20} use the same model to estimate the cost and cardinality 
and Neo~\cite{CM_L_MarcusNMZAKPT19} uses the latency (performance) of (sub)plan as the cost. A model, which is able to capture the data itself, operator level information, and subplan information, can predict the cost accurately. For example, the work~\cite{CE_L_SP_SunL20}, one of the state-of-the-art methods,
 adopts the tree-LSTM model to capture the information mentioned above.
However, all of them are supervised methods. If the workload shifts or the data is updated the models need to be retrained from the scratch. 

\subsubsection{Possible Future Directions} There are two possible directions as follows:
\begin{enumerate}[noitemsep]
\item \textbf{Cloud database systems.} The users of the cloud database systems need to meet their latency or throughput at the lowest price.
Integrating the economic cost of running queries into the cost model is a possible direction. It is interesting to consider these related information into the cost model.
For example, \citet{CM_L_SiddiquiJQPL20} consider the number of container into their cost model.  
\item \textbf{Learning-based methods.} Learning-based methods to estimate the cost also suffer from the same problems with methods in cardinality estimation (Section \ref{PFD_CE}). 
The model that has been adopted in a real system is a light model~\cite{CM_L_SiddiquiJQPL20}.
The trade-off between accuracy and training time is still a problem. The possible solutions adopted in cardinality estimation also can be used in the cost model.
\end{enumerate}

\section{Plan Enumeration}\label{PE}
In this section, we present the researches published to handle the problems in plan enumeration. We classify the work on plan enumeration into two groups, non-learning methods and learning-based methods.
\subsection{Non-learning Methods}
In 1997, \citet{JOIN_T_TD_SteinbrunnMK97} proposed a representative survey for selecting an optimal join orders. Thus, we mainly focus on the researches after 1997.
\subsubsection{Dynamic Programming}
\citet{JOIN_T_SelingerACLP79} propose a dynamic programming algorithm to select the optimal join order for a given conjunctive query. 
They generate the plan in the order of increasing size and restrict the search space to left-deep trees, which significantly speeds up the optimization.
\citet{JOIN_T_VanceM96} propose a dynamic programming  algorithm to find the optimal join order by considering different partial table sets. 
They use it to generate the optimal bushy tree join trees containing cross products.
~\cite{JOIN_T_SelingerACLP79, JOIN_T_VanceM96} are generate-and-test paradigm and most of the operations are used to check whether the subgraphs are connected and two subgraphs 
are combinative. Thus, none of them meet the lower bound in~\cite{JOIN_T_OnoL90}.
\citet{JOIN_T_TD_MoerkotteN06} propose a graph-based Dynamic programming algorithm. They first introduce a graph-based method to generate the connected subgraph. 
Thus, it does not need to check out the connection and combinations and perform more efficiently. Then, they adopt DP over them for the generation of optimal bushy tree without cross products.
\citet{JOIN_T_MoerkotteN08} extend the method in~\cite{JOIN_T_TD_MoerkotteN06} to deal with non-inner joins and a more generalized graph, hyper graph, where join predicates can involve more than two relations.

\subsubsection{Top-down Strategies} TDMinCutLazy is the first efficient top-down join enumeration algorithm proposed by \citet{JOIN_T_TD_DeHaanT07}.
They utilize the idea of minimal cuts to partition a join graph and introduce two different pruning strategies, predicted cost bounding and accumulated cost bounding 
into top-down partitioning search, which can avoid exhaustive enumeration. 
Top-down method is almost as efficient as dynamic programming and has other tempting properties, e.g., pruning, interesting order.
\citet{JOIN_T_TD_FenderM11} propose an alternative top-down
join enumeration strategy (TDMinCutBranch). TDMinCutBranch introduces
a graph-based enumeration strategy which only generates the
valid join, i.e. cross-product free partitions, unlike TDMinCutLazy which adopts a generate-and-test approach.
In the following year, \citet{JOIN_T_TD_FenderMNL12} propose another
top-down enumeration strategy TDMinCutConservative
which is easier to implement and gives better runtime performance in comparison to TDMinCutBranch.
Furthermore, \citet{JOIN_T_TD_FenderM13,JOIN_T_FenderM13} present a general framework to handle non-inner joins and a more generalized graph, hyper graph for top-down join enumeration.
RS-Graph, a new join transformation rules based on top-down join enumeration, is presented in~\cite{PE_T_ShanbhagS14} to efficiently generate the space of cross-product free join trees.
\subsubsection{Large Queries} For large queries, Greedy Operator Ordering~\cite{JOIN_T_Fegaras98} builds the bushy join trees bottom-up by adding the most profitable (with the smallest intermediate join size) joins first.
To support large queries, \citet{JOIN_T_KossmannS00} propose two different incremental dynamic programming methods, IDP-1 and IDP-2.
With a given size k, IDP-1 runs the algorithm in \cite{JOIN_T_SelingerACLP79} to construct the cheapest plan with that size and then regards it as a base relation, and repeats the process. 
With a given size k, in every iteration, IDP-2 first performs a greedy algorithm to construct the join tree with k tables and 
then runs DP on the generated join tree to produce the optimal plan, regards the plan as a base relation,
and then repeats the process. 
\citet{JOIN_T_Neumann09} proposes a two-stage algorithm: First, it performs the query simplification to restrict the query graph by the greedy heuristic 
until the graph becomes tractable for DP, and then it runs a DP algorithm to find the optimal join orders for the simplified the join graph.
\citet{JOIN_T_BrunoGJ10} introduce enumerate-rank-merge framework which generalizes and extends the previous heuristics~\cite{JOIN_T_Fegaras98,JOIN_T_Swami89}. 
The enumeration step considers the bushy trees. The ranking step is used to evaluate the best join pair each step, which adopts the min-size metric. The merging step constructs the selected join pair.
\citet{JOIN_T_TD_NeumannR18} divide the queries into three types: small queries, medium, and large queries according to their query graph type and the number of tables. 
Then, they adopt DP to solve small queries, restrict the DP by linearizing the search space for medium queries, and use the idea in~\cite{JOIN_T_KossmannS00} for large queries.

\subsubsection{Others}
\citet{JOIN_T_Trummer017} transform the join ordering problem into a mixed integer linear program to minimize the cost of the plan and adopt the existing the MILP solvers to obtain a linear join tree (left-deep tree). 
To satisfy the linear properties in MILP, they approximate the cost of scan and join operations via linear functions.

Most of existed OLAP systems mainly focus on start/snowfake join queries (PK-FK join relation) and generate the left-deep binary join tree. 
When handling FK-FK join, like in TPC-DS (snowstorm schema), they inccur a large number of intermediate results. 
\citet{JOIN_T_TD_NamH020} introduce a new n-ary join operator, which extends the plan space. 
They define the core graph to represent the FK-FK joins in the join graph and adopt the n-ary join operator 
to process it. They design a new cost model for this operator and integrate it into an existed  OLAP systems.
\subsection{Learning-based Methods}
\begin{figure}[t]
    \centering
    \includegraphics[width=\linewidth]{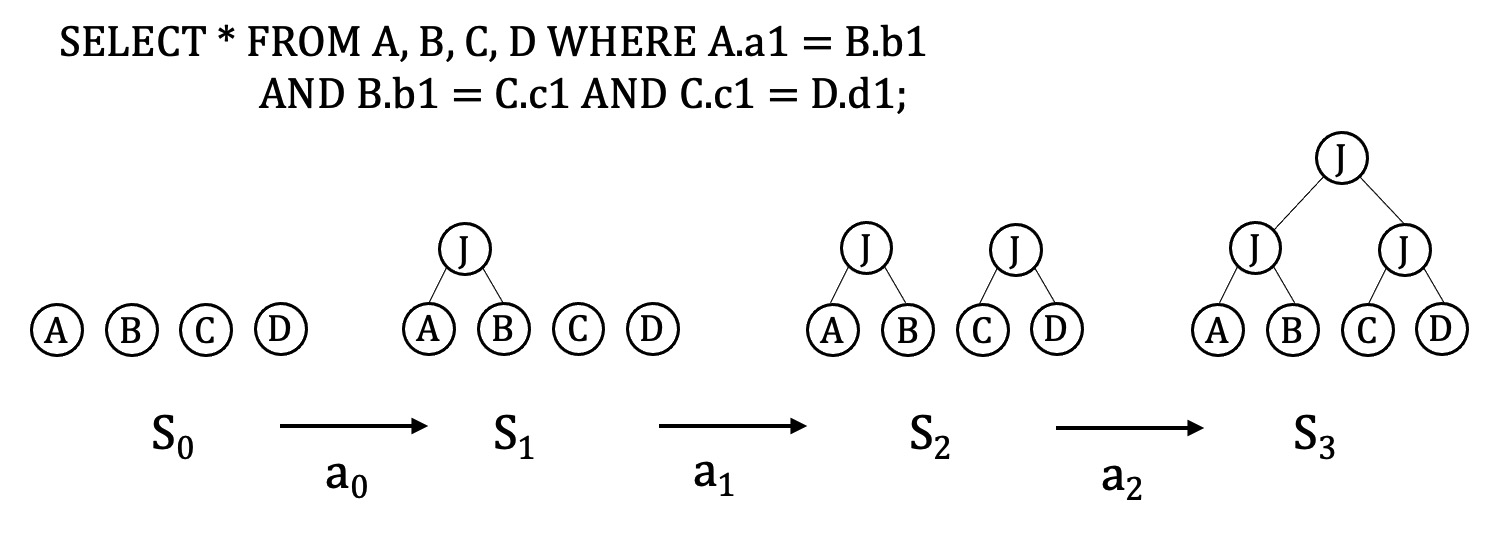}
    \caption{One possible join order episode}
    \Description{RL methods}
    \label{join_order_rl}
  \end{figure}
All learning-based methods adopt the reinforcement learning (DL).
In RL, an agent interacts with environment by actions and rewards. At each step \(t\), the agent uses
a policy \(\pi\) to choose an action \(a_t\) according to the current state \(s_t\) and transitions to 
a new state \(s_{t+1}\). Then the environment applies the action \(a_t\) and returns a reward \(r_t\) to the 
agent. The goal of RL is to learn a policy \(\pi\), a function that automatically takes an action 
based on the current state, with the maximum long-term reward. In join order selection, state is the current sub-trees, 
and action is to combine two sub-trees, like in Figure \ref{join_order_rl}. The reward of intermediate action is 0 and
the reward of the last action is the cost or latency of the query.

ReJoin~\cite{PE_L_MarcusP18} adopts the deep reinforcement learning (DRL), which has widely been adopted in other areas, e.g., influence maximization~\cite{DSE_rl1}, to identify the optimal join orders. 
State in DRL represent the current subtrees. Each action will combine two subtrees together into a single tree. It uses cost obtained from the cost model in optimizer as the reward. 
ReJoin encodes the tree structure of (sub)plan, join predicates, and selection predicates in state.
Different with~\cite{PE_L_MarcusP18}, \citet{PE_L_abs-1911-11689} create a matrix to represent a table or a 
subquery in each row and adopt the cost model in~\cite{Leis2015} to quickly obtain the cost of one query.
DQ~\cite{PE_T_abs-1808-03196} is also a DRL-based method. It uses one-hot vectors to encode the visible attributes in the (sub)query. 
DQ also encodes the choice of physical operator by adding another one-hot vector.
When training the model, DQ first uses the cost observed from the cost model of the optimizer and then fine-tunes the model with true running time.
\citet{PE_L_Yu0C020} adopt DRL and tree-LSTM together for join order selection. Different with the previous methods~\cite{PE_L_MarcusP18,PE_L_abs-1911-11689,PE_T_abs-1808-03196}, tree-LSTM
can capture more the structure information of the query tree. Similar with~\cite{PE_T_abs-1808-03196}, they also use cost to train the model and then switch to
running time as feedback for fine-tuning. Notice, they also discuss how to handle the changes in the database schema, e.g. adding/deleting the tables/columns.
SkinnerDB~\cite{PE_L_TrummerWMMJA19} adopts the UCT, a reinforcement learning algorithm, and learns from the current query, while the previous learning-based join order methods are learning from previous queries.
It divides the execution of one query into many small slices where different small slices may choose the different join order and learn from the previous execution slices.
\subsection{Our Insights}
\subsubsection{Summaries}
The non-learning based studies focus on improving the efficiency and the ability (to handle the more general join cases) of the existing approaches. 
Compared with dynamic programming approach, the top-down strategy is tempting due to the better extensibility, 
e.g., adding new transformation rules, branch-and-bound pruning.
Both of them have been implemented in many database systems.

Compared with the non-leaning methods, learning-based approaches have a fast planning time. 
All learning-based methods employ reinforcement learning. The main differences between them are:(1) choosing which information as the state and how to encode them, (2) adopting which models.
A more complicated model with more related information can achieve better performance. The state-of-the-art method~\cite{PE_L_Yu0C020} adopts a tree-LSTM model similar with~\cite{CE_L_SP_SunL20} to generate the representation of a subplan.
Due to the inaccuracy in the cost model, it can improve the quality of model by using the latency to fine-tune the model. 
Although current state-of-the-art method~\cite{PE_L_Yu0C020} outperforms the non-learning based methods as shown in their experiments, how to integrate the learning-based method into the real system must be solved. 

\subsubsection{Possible Future Directions} There are two possible directions as follows:
\begin{enumerate}[noitemsep]
  \item \textbf{Handle large queries.} All methods proposed to handle large queries are DP-based methods in the bottom-up manner.
  A question is remaining: how to make the top-down search strategy support the large queries. Besides, the state-of-art method for large queries~\cite{JOIN_T_TD_NeumannR18}
  cannot support the general join cases.
  \item \textbf{Learning-based methods.} Current leaned methods only focus on the PK-FK join and the join type is inner join. How to handle the other join cases is a possible direction.
  None of the proposed methods have discussed how to integrate them into a real system. In their experiments, they implement the method as a separate component
  to get the right join order, and then send to the database. The database still need to optimize it to get the final physical plan. If a query has subquery, they may 
  interact multiple times. The reinforcement learning methods are trained in a certain environment, which refers to a certain database in the join order selection problem.
  How to handle the changes in the table schemas and data is also a possible direction.
\end{enumerate}

\section{Conclusion} \label{RS}
Cardinality estimation, cost model, and plan enumeration play critical roles to generate an optimal execution
plan in a cost-based optimizer.
In this paper, we review the work proposed to improve their qualities, 
including the traditional and learning-based methods. 
Besides, we provide possible future directions respectively.

We observe that more and more learning-based methods are introduced and outperform traditional methods. 
However, they suffer from long training and updating time. 
How to make the models robust to workload shifts and data changes or to update models quickly is still an open question. 
Traditional methods with theoretical guarantees are widely adopted in real systems.
There is a great possibility of improving traditional methods with new algorithms and data structures.
Moreover, We believe the ideas behind the traditional methods can be used to enhance the learning-based methods.

\begin{acks}
  Zhifeng Bao is supported in part by ARC DP200102611, DP180102050, and a Google Faculty Award.
\end{acks}
\bibliographystyle{ACM-Reference-Format}
\bibliography{ms}
\end{document}